\documentclass{aastex}
\usepackage{spr-astr-addons}
\usepackage{url}

\RequirePackage{color}

\newcommand{\emaila}{avdan.hsn@gmail.com}

\begin{document}

\title{A search for periodicities from a ULX in the LINER galaxy NGC 4736}
\shorttitle{A ULX in the LINER Galaxy NGC 4736}
\shortauthors{Avdan et al.}

\author{Hasan Avdan} 
\and
\author{Senay Kayaci Avdan} 
\and
\author{Aysun Akyuz}
\affil{Department of Physics, University of Cukurova, Adana, Turkey}
          \and
\author{Solen Balman}
\affil{Department of Physics, Middle East Technical University, Ankara, Turkey}
\email{\emaila}

\begin{abstract}

We report our findings on a new quasi-periodic oscillation (QPO) and a long period from the ultraluminous X-ray source (ULX) X-2 in nearby galaxy NGC 4736 based on the {\it Chandra} and {\it XMM-Newton} archival data. To examine the timing properties, power density spectra of the source have been obtained using Fast Fourier Transform. Also the spectral parameters of the source have been calculated by obtaining and fitting the energy spectra. Power density spectrum of this source reveals a QPO peak at $0.73_{-0.14}^{+0.16}$ mHz with an fractional rms 
variability of $16\%$ using the {\it Chandra} data (in the year 2000-lower state of the source). 
The {\it XMM-Newton} data analysis indicates a peak at 
$0.53_{-0.35}^{+0.09}$ mHz with a fractional rms variation of $5\%$ (in the year 2006-higher state of the source). 
These recovered QPOs overlap within errors and may be the same oscillation.
In addition, we detect a long periodicity or a QPO in the {\it Chandra} data of about $(5.2\pm2.0)\times10^{-5}$ Hz ($\sim$ 5.4 hrs) 
over 3 $\sigma$ confidence level. If this is a QPO, it is the lowest QPO detected from a ULX. 
The mass of the compact object in ULX X-2 is estimated using the Eddington luminosity and a disk blackbody model in the range (10$-$80) M$_{\sun}$.

\end{abstract}


\section{Introduction}

Ultraluminous X-ray sources (ULXs) are off-nuclear point-like sources that have X-ray luminosities ($L_{x} \textgreater 3 \times 10^{39}$ erg $s^{-1}$) above the Eddington limit of a 20$M_{\sun}$ black hole (BH) (Feng \& Soria 2011). Although the true nature of these objects 
is not clear, current models propose several alternatives to explain their high luminosities: It could 
either be due to the geometric beaming if ULXs are powered by an accreting stellar black hole (King et al. 2001) 
or the super-Eddington fluxes originating from disks (Begelman 2002). Additionally, some ULXs could be powered 
by intermediate-mass black holes (Miller \& Colbert 2004). 

X-ray spectral and temporal analyses allow us to study the nature of ULXs and to constrain BH masses in ULXs 
(Soria \& Ghosh 2009). Quasi-Periodic Oscillations (QPOs) in X-ray binaries provide information about 
the inner accretion disk structure around the compact object (Mucciarelli et al. 2006). Up to now, 
QPOs have been detected from a few ULXs in nearby galaxies. The first ULX (X-1) that showed strong evidence 
for a QPO was in M82 (Strohmayer \& Mushotzky 2003). Using a 30 ks {\it XMM-Newton} observation they found 
that the power-density spectrum (PDS) of this source shows a prominent QPO peak at a frequency of 54 mHz with 
an integrated rms of 8.5\% in the (2$-$10) keV. They also determined a 107 mHz QPO from the same ULX using 
archived {\it RXTE} observations. Based on the Schwarzschild geometry, they calculated the mass of the BH 
in source X-1 as \textless 1.87$\times10^{4}$ $M_{\sun}$ using the highest QPO frequency. In the other 
studies using the {\it XMM-Newton} and {\it RXTE} observations similar properties were detected 
for the ULX in M82 (Mucciarelli et al. 2006, Feng \& Kaaret 2007). Also, Feng et al. (2010) reported QPOs at (3$-$4) mHz from a transient ULX (X42.3+59) in M82 using three {\it Chandra} and two {\it XMM-Newton} observations. They estimated the BH mass in the range of (1.2$-$4.3$)\times10^{4}$ $M_{\sun}$ in the ULX by scaling the QPO frequency to that of their type (A/B) of QPOs in stellar mass BHs. Liu et al. (2005) have presented the PDS 
of a ULX in NGC 628 (M74) showing a broad peak at a frequency range of (0.1$-$0.4) mHz in one 
{\it XMM-Newton} and two {\it Chandra} observations with an rms variation of 13.8\% and 23.9\% in 
(2$-$4) keV and (4$-$10) keV band, respectively. They estimated the mass of the compact object as 
$\sim$(2$-$20)$\times10^{3}$ $M_{\sun}$ using the scaling relation between break frequency and BH masses. 
Another ULX (Holmberg IX X-1) which showed a QPO was detected in Holmberg IX, using a 119 ks {\it XMM-Newton} observation 
(Dewangan et al. 2006b). It has a centroid frequency of 202.5 mHz with an rms of 6\% in the (0.2$-$10) keV 
energy band. However, Heil et al. (2009) reported that while the source Holmberg IX X-1 shows variability, it does not show a significant QPO feature. Strohmayer et al. (2007) found a ULX in NGC 5408 (X-1) which shows a pair of QPOs with a 4:3 ratio. 
The first peak is at 20 mHz and the second peak is at 15 mHz, at integrated rms of 9\% from a 130 ks 
{\it XMM-Newton} observation. Considering this QPO is analogous to the ones in Galactic systems (for example GRO J1655-40), 
they found a BH mass range of (1.5$-$3.5)$\times10^{3}$ $M_{\sun}$. 
Furthermore, using the mass-disk temperature scaling, $kT_{disk} \propto M^{-1/4}$, they also derived another mass range (1.81$-$4.74)$\times10^{3}$ $M_{\sun}$. Pasham and Strohmayer (2012) also presented detailed study of NGC 5408 X-1 using recent {\it XMM-Newton} observations. They detected QPOs in the range of (10$-$40) mHz with new observations. They also calculated a lower limit of $\sim800M_{\sun}$ on the mass of the BH in X-1 by scaling the minimum QPO frequency to a transition frequency of a reference stellar BH with known mass. Another ULX in nearby galaxy NGC 6946 shows possible QPO feature with a central frequency of $\sim$8.5 mHz (Rao et al. 2010). They calculated an integrated rms amplitude of 59\% for X-1 in the (1$-$10) keV energy range and a BH mass of $\sim 10^{3}$ $M_{\sun}$ for the compact source by scaling the frequency with mass.  

In the present work, we search the X-ray timing variations and spectral properties of the ULX X-2 in the 
LINER galaxy NGC 4736. The source is located in the disk of the galaxy. Akyuz et al. (2013) examined the 
long term light curves and energy spectrum of this source using {\it XMM-Newton} archival data revealing its 
transient nature. Previously, this source was detected as a point source using the {\it Chandra} archival data (Liu 2011) with no reference to its transient nature.
A more recent study by Lin et al. (2013), also discusses the nature of the ULX X-2 using the {\it ROSAT}, {\it XMM-Newton} and {\it Chandra} archival data. They show highly variable 
dipping behaviour of the light curve in the brightest observations and discuss spectral properties in the dipping periods. 
They also examine the {\it HST} images and identify a point-like red optical counterpart candidate. Considering the colors 
and the luminosities of this candidate, they suggest that it could be a G8 supergiant or a dwarf star in the globular cluster.

Differently from the Lin et al. (2013) work, we concentrate on the timing properties and search for periodicities from the ULX X-2. This paper organized as follows: The observations and methods used in data reductions are described in Sect. 2. The timing and spectral analysis are given in Sect. 3 and Sect. 4. Discussions on the QPOs and the BH mass estimations are given in Sect. 5.

\begin{figure}[t]
  \includegraphics[scale=0.46, angle=0]{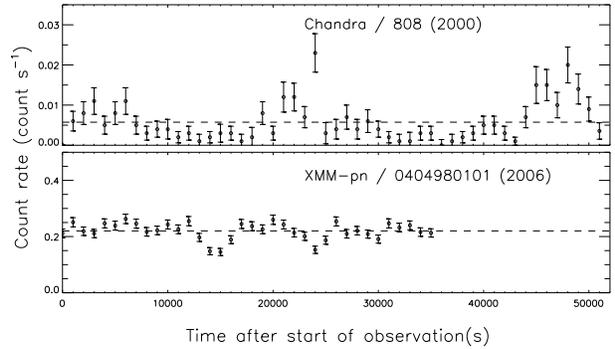}
    \caption{The X-ray light curves of X-2 in the (0.2$-$10) keV energy band. Observation dates are labelled on the light curves. The dashed lines show the mean count rate. High background flare times are excluded from the ObsID 0404980101. Both light curves have a bin size of 1000 s.}
 
\end{figure}

\section{Observations And Data Reduction}

In this study we used the {\it XMM-Newton} and the {\it Chandra} 
archival observations which have the longest exposure for the analysis of X-2.
 The {\it XMM-Newton} observation was carried out on 2006 November 27 for 55 ks (ObsID 0404980101) and the {\it Chandra} ACIS-S observation was carried out on 2000 May 13 for 49.8 ks (ObsID 808). We excluded high background flarings from ObsID 0404980101 data set, which removed the last $\sim$ 17 ks of the observation. Additionally, we did search the shorter exposures ({\it XMM-Newton} ObsID 0094360701, {\it Chandra} ObsID 9553) as well however, there was no indication of significant QPOs or periodicities at any frequency.
 
Data reductions were performed using the SAS (Science Analysis Software) version 12.0 for {\it XMM-Newton} and the CIAO (Chandra Interactive Analysis of Observations) version 4.3 with the CALDB (Calibration Database) version 4.4.2 for {\it Chandra}. In {\it XMM-Newton}, the events corresponding to PATTERN$\leq$ 4 were selected with FLAG==0 option for the pn camera, PATTERN$\leq$ 12 were used for the MOS cameras. The events were extracted from a circular region of $18^{\arcsec}$ enclosing the centroid position of the source (R.A. = $12^{h}$ $50^{m}$ $48^{s}$.6, Dec = $41^{\circ}$ $07^{\arcmin}$ $43^{\arcsec}$) in {\it XMM-Newton}. In the {\it Chandra} observation, X-2 was located in the ACIS-S3 (back-illuminated) chip. The events were extracted from a circular region of $2^{\arcsec}$ surrounding the centroid position of the source (R.A. = $12^{h}$ $50^{m}$ $48^{s}$.6, Dec = $41^{\circ}$ $07^{\arcmin}$ $42^{\arcsec}$.5) in {\it Chandra}. Background photons were extracted using a proper region from a location with no source contamination. Figure 1 shows the two background-subtracted light curves with 1000 s time bins for display purposes only. The light curves were calculated using {\it lcmath} and {\it lcurve} tasks within Xronos version 5.21 for the {\it XMM-Newton} pn and {\it Chandra} data. The same data used for timing analysis by extracting light curves of much higher time resolution is described in Sect. 3.

\begin{figure*}
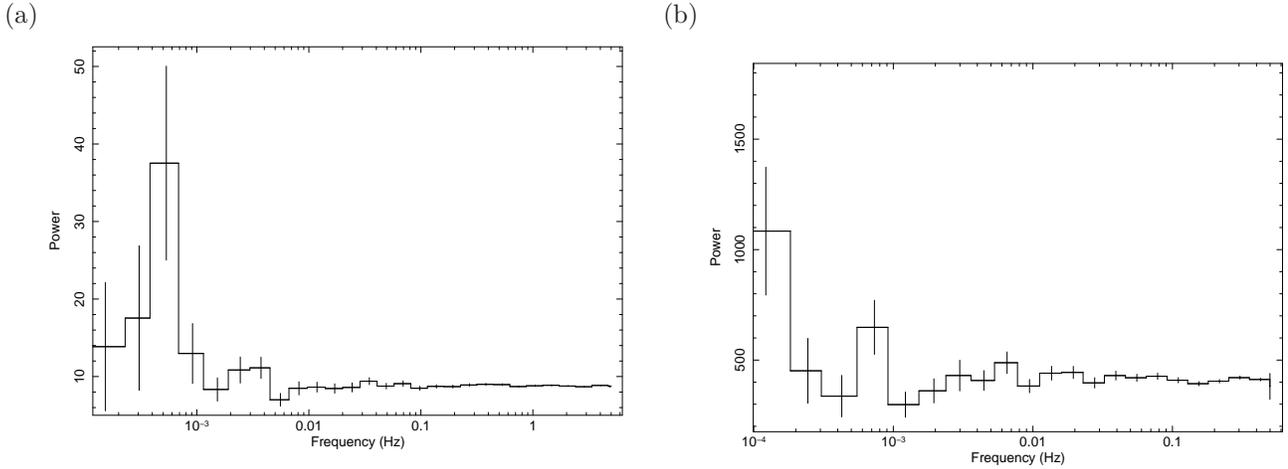

(a)
{
\label{fig:sub:a}
\includegraphics[scale=0.33, angle=-90]{avdan_fig2.ps}
}
(b)
{
\label{fig:sub:b}
\includegraphics[scale=0.33, angle=-90]{avdan_fig3.ps}
\caption{Averaged power spectrum of X-2; (a) obtained using the {\it XMM-Newton}-pn observation (ObsID 0404980101), (b) obtained using the {\it Chandra} ACIS-S observation (ObsID 808). We used the Miyamoto normalization for power in units of (rms/mean)$^2$/Hz (Miyamoto et al. 1991).}
}
\end{figure*}

\begin{figure}

\label{fig}
\includegraphics[scale=0.30, angle=-90]{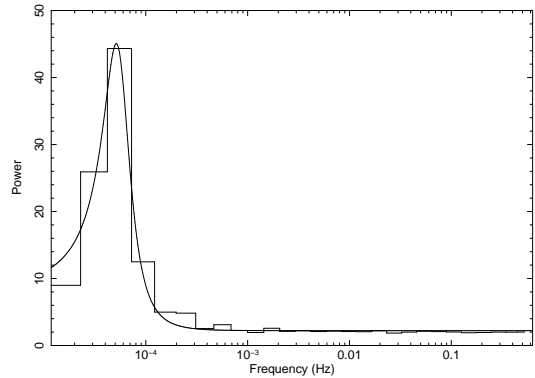}
\caption{Power density spectrum of X-2 obtained using the {\it Chandra} ACIS-S observation (ObsID 808). The PDS is 
normalized according to Leahy et al. (1983).}
\end{figure}

\section{Timing Analysis}

For the {\it XMM-Newton} observation we used only the pn data (ObsID 0404980101) in timing analysis. 
To search the quasi periodicity for X-2 the light curve was sampled at 0.1 s for the PDS. A 2nd-order polynomial trend was removed from the light curve to clean the red noise in lower frequencies. Then the resultant light curve was divided into six intervals. The average PDS was obtained by calculating and averaging the power spectra for each interval using Xronos version 5.21. The PDS were normalized according to Miyamoto et al. (1991). 
Figure 2a shows the PDS in the (0.2$-$10) keV energy band and shows a QPO peak around at $\sim0.5$ mHz 
(2.4 $\sigma$) with an integrated rms of $5\% \pm 1\%$. The significance is calculated assuming SNR (in $\sigma$) $=(P_{m} - P_{ref})/E_{p}$ using the peak value ($P_{m}$), the continuum value ($P_{ref}$) and error of the peak value ($E_{p}$) (Balman 2010 and references therein). When determining the significance, the continuum value is calculated in the (0.1 $-$ 2) mHz frequency range. To clarify that this is a real peak and not an instrumental or dither related artifact, 
we calculated the PDS of the other sources in the field close by as well as the background and confirmed that there are no similar peaks. Additionally, we carefully repeated the PDS analysis taking care of existing data gaps in the {\it XMM-Newton} data. In order to do this we derived light curve for each GTI and calculated the PDS by combining them. We were not able to identify any fake QPO peaks that would result from such gaps. 
We fitted the {\it XMM-Newton} PDS using a constant frequency model for the flat noise component plus a Lorentzian component for the QPO. Our best fitting composite model has a QPO centroid frequency of $\nu_{QPO}$ = $0.53_{-0.35}^{+0.09}$ mHz with a width of ${\sigma}_{FWHM}$ = $0.10$ mHz. 
This yields $Q = {\nu}_{QPO}/{\sigma}_{FWHM} = 5.3$. The constant model component which represents the continuum value in the PDS has a value of $\sim$9 (rms/mean)$^2$/Hz. Moreover, 
we obtained the Leahy-normalized PDS (Leahy et al. 1983) for comparison and we found the same frequency value 
and significance level for the QPO.

We also did the same analysis for the {\it Chandra} (ObsID 808) data. The light curve was sampled at 1 s for the PDS (frame time of this observation is 0.84104 s). The resultant time series were divided into seven intervals and the individual PDS were averaged. Figure 2b shows the PDS for the {\it Chandra} in the (0.2$-$10) keV range. We find a QPO peak around at $\sim0.7$ mHz (2.5 $\sigma$) with an integrated rms of $16\% \pm 3\%$. The significance is calculated in the same manner as in the
{\it XMM-Newton} analysis, except the continuum value is calculated in the (0.2$-$2) mHz frequency range for this data.
We used the same method as in the {\it XMM-Newton} analysis to rule out
any dither, observational or instrumental related peaks from our detected QPO of which we did not find any. 
We fitted the PDS using the same composite model (Lorentzian plus a constant). Our best fitting model has a QPO centroid frequency of $\nu_{QPO}$ = $0.73_{-0.14}^{+0.16}$ 
mHz with a width of ${\sigma}_{FWHM}$ = $0.10$ mHz ($Q = 7.3$) and a constant model value of $\sim$336 (rms/mean)$^2$/Hz. 

In addition, we detected a possible periodic oscillation or another QPO around $\sim5.2\times10^{-5}$ Hz ($\sim$5.4 hrs)
above 3 $\sigma$ confidence level using the {\it Chandra} 
observation (ObsID 808). Figure 3 shows the PDS obtained from a single Fast Fourier Transform (FFT). The PDS is normalized 
according to Leahy et al. (1983) and the three sigma detection level is above a power of 32 (calculated according to van der Klis
1989). We fitted the PDS using a Lorentzian plus a constant model. The best fitting model has a centroid frequency of 
$\nu$ = $5.2\times10^{-5}$ Hz with a width of ${\sigma}_{FWHM}$ = $0.42\times10^{-4}$ Hz and a constant noise level of 
$\sim$2.22. The 50 ks time span of the data can be used to calculate $\Delta\nu$ for the detected frequency yielding $(5.2\pm2.0)\times10^{-5}$ Hz. We could not look for this frequency in the {\it XMM-Newton} data since the removal of flares reduced the exposure/time span and such low frequencies were not revealed.  

\section{Spectral Analysis}

{\it XMM-Newton} (ObsID 0404980101) and {\it Chandra} (ObsID 808) observations provide sufficient statistics for spectral analysis. To improve statistics for {\it XMM-Newton} analysis both pn and MOS data were used. All spectra were grouped to at least 20 counts per bin and analysed using XSPEC version 12.5. 

Spectral analyses reveal that the source spectrum is best fitted by a composite model of a power-law plus a disk blackbody 
(PL+DISKBB) (${\Gamma}$ $\sim$ 1.7, $T_{in}$=0.75 keV) using the {\it XMM-Newton} data and one-component model of a power-law (PL) (${\Gamma}$ $\sim 2.5$) using the {\it Chandra} data. 
The best fitted model parameters with the {\it XMM-Newton} pn, MOS and the {\it Chandra} ACIS-S 
data are given in Table 1. Model parameters of {\it XMM-Newton} data are taken from Akyuz et al. (2013). They already performed a more detailed spectral modelling and short/long-term timing variability. As a result of these analyses they conclude that X-2 possesses a transient nature. In the table, the flux values represent 
the unabsorbed fluxes and have been calculated for (0.3$-$10) keV band. The source is $\sim$20 times brighter 
in {\it XMM-Newton} than in {\it Chandra} observation, strongly indicating time variability. {\it Chandra} energy spectrum of X-2 is given in Figure 4. {\it XMM-Newton} energy spectrum of this source can be seen in Figure 11 of Akyuz et al. (2013). 

\begin{figure}

\label{fig}
\includegraphics[scale=0.30, angle=-90]{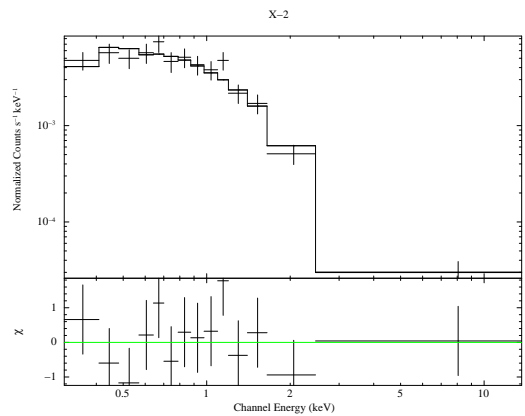}
\caption{Energy spectrum of X-2 obtained using the {\it Chandra} ACIS-S observation (ObsID 808).}
\end{figure}

\begin{table*}
\small
\caption{Spectral Model Parameters For NGC 4736 X-2}
\begin{tabular}{c c c c c c c c c}
\tableline

 & $N_{H}$($10^{22}$) & $\Gamma$  & kT   & K$_{PL}$\tablenotemark{b}  &  
K$_{D}$\tablenotemark{c} & $\chi^{2}$/dof  & F\tablenotemark{d}($10^{-13}$)    & L\tablenotemark{e}($10^{38}$)  \\

&  $cm^{-2}$ & keV &   & ($10^{-5}$) & & & erg $cm^{-2}$ $s^{-1}$ & erg $s^{-1}$ \\

\tableline

& & & & & & & &\\

{\it XMM-Newton}\tablenotemark{a} & $0.02_{-0.01}^{+0.01}$ & $1.72_{-0.13}^{+0.14}$ & $ 0.75_{-0.04}^{+0.06}$ &  $5.79_{-1.10}^{+1.40}$  & $0.05_{-0.01}^{+0.02}$  &  530.81/517 & 7.57 & 16.73   \\

& & & & & & & & \\

{\it Chandra} & $0.07_{-0.03}^{+0.06}$  & $ 2.55_{-0.26}^{+0.38}$ & - & $0.77_{-0.18}^{+0.28}$ & - & 8.27/11 & 0.37 & 0.85     \\

& & & & & & &\\
\tableline
 
\tableline                                           
\end{tabular}
\tablenotetext{a}{{\it XMM-Newton} model parameters were taken from Akyuz et al. (2013).}
\tablenotetext{b}{Normalization parameter of the PL model in units of photon cm$^{-2}$ s$^{-1}$ keV$^{-1}$ at 1 keV.}
\tablenotetext{c}{Normalization parameter of the DISKBB model. K$_{D}$ = $[(R_{in}/km)/(D/10 kpc)]^{2}\times \cos\theta$, where R$_{in}$ is the inner disk radius, D is the distance to the source and $\theta$ is the angle of the disk.}
\tablenotetext{d}{Unabsorbed flux in the (0.3$-$10) keV energy band.}
\tablenotetext{e}{These values have been calculated using a distance of 4.3 Mpc (Winter et al. 2006).}
\tablecomments{Spectral models are PL+DISKBB for {\it XMM-Newton} and PL for {\it Chandra}.}
\end{table*}

\section{Discussion And Summary}

In this work, we have presented the energy and power spectra of the ULX X-2 in NGC 4736 
using archival {\it Chandra} and {\it XMM-Newton} observations.

{\it Chandra} observation of the source 
revealed a QPO with a centroid frequency of $\nu_{QPO}=0.73_{-0.14}^{+0.16}$ mHz (with an integrated rms fractional 
variability of $16\% \pm 3\%$ and $Q = 7.3$) while the {\it XMM-Newton} observation 
showed a QPO at $0.53_{-0.35}^{+0.09}$ mHz (with an integrated rms 
of $5\% \pm 1\%$ and a $Q = 5.3$). The two frequencies overlap within their error limits, thus these could be
either slightly differing QPOs or the same QPO from the system. 
It is consistent that the low frequency QPO has diminished in rms percent variability when the source changed 
to a higher luminosity state with the outer parts of the disk being more stable to variations. It is expected
that in high states of BH LMXBs low frequency QPOs are suppressed.

The energy spectrum for the {\it Chandra} observation shows a best fit with a power-law model (${\Gamma}$ $\sim 2.5$). 
The {\it XMM-Newton} spectrum is best fitted with a two-component model, the disk blackbody plus power-law ($T_{in}$=0.7 keV, ${\Gamma}$ $\sim$ 1.7) as published earlier by Akyuz et al. (2013). Our analysis showed that the source luminosity has changed by a factor $\sim$20 times between two observations. There are also couple of transient ULXs that exhibits decrease in luminosity below the ULX regime: CXOM31 J004253.1+411422 in M31 (Kaur et al. 2012), ULX in M83 (Soria et al. 2012), 1RXH J132519.8-430312 and CXOU J132518.2-430304 in NGC 5128 (Burke et al. 2013). The level of change in luminosity of the source is not well-known since the source was observed only four times within eight years. Its spectral shape has changed from a one-component model of a PL to a 
two-component model of PL+DISKBB where the disk component contributed $\sim60\%$ and the power-law component yielded the rest of the total flux. This indicates that X-2 may be in a spectral state analogous to the thermal state of Galactic BH X-ray binaries (Remillard \& McClintock 2006) during the {\it XMM-Newton} observation. The source has a steep PL photon index as calculated from the {\it Chandra} observation with a lower luminosity than the thermal state. However, the steep power law state of Galactic BH X-ray binaries is characterized by a relatively high luminosity with a PL component of ${\Gamma} > 2.4$ (Remillard \& McClintock 2006). The ULX X-2 exhibits opposite behaviour, however a similar unusual low/soft state has also been observed from ULX X-2 in NGC 1313 (Feng \& Kaaret 2006). This unusual behaviour may arise from the lack of good statistics in the {\it Chandra} data. On the other hand, Lin et al. (2013) also carried out spectral analysis of X-2 using one {\it ROSAT}, one {\it XMM-Newton} and two {\it Chandra} observations. Their best-fitting spectral models, PL model in the {\it Chandra} (ObsID 808) and PL+MCD model in the {\it XMM-Newton} (ObsID 0404980101) data, and the derived spectral parameters are in agreement with ours. They noticed that the source might be in the hard state instead of the steep power-law, if one takes into account the large uncertainty of PL photon index ($\sim 2.5 \pm0.5$) due to the poor quality of the data.

The mass of the BH in ULX systems can be calculated using the relation 
M=$c^{3}/(2\pi 6^{3/2}G {\nu}_{QPO})$ $\simeq$ 2190/${\nu}_{QPO}$ $M_{\sun}$ (van der Klis 2006). 
This is based on the assumption that the QPO frequencies are associated with the Keplerian frequency 
at the innermost circular orbit around a Schwarzschild BH. The formula yields a mass value of
$\sim$ $3.5\times 10^{6}$ $M_{\sun}$ for X-2, which is rather high for even an intermediate BH. Another approach 
is to consider the inverse proportionality between the BH mass and the QPO frequencies, and how this scales over different systems. 
Dewangan et al. (2006a) 
used this proportionality for ULX X-1 in M82 and found the mass of the BH to be in the range of (25$-$520)$M_{\sun}$. 
We used the same scaling argument to approximate the mass of the BH in X-2 of NGC 4736. We chose the ULX in 
NGC 628 for the scaling since the detected QPOs have similar low frequencies and luminosities ($L_{x}=(4.5-13.4)\times10^{38}$ erg s$^{-1}$, Liu et al. 2005). Also, this source is one of the ULX which have the lowest QPO frequency ever detected. Using this scaling argument and assuming the mass ($(2-20)\times 10^{3}$ $M_{\sun}$) and QPO frequencies ($(1-4)\times 10^{-4}$ Hz) of the ULX in NGC 628 (Liu et al. 2005), we estimated a BH mass of $(2-400)\times 10^{2}$ $M_{\sun}$ for the compact source in X-2. However, we note that some ULXs have low luminosities and they were estimated to have stellar/massive-stellar ($10 M_{\sun} \leq M \leq 100M_{\sun}$) BHs (Kaur et al. 2012; Soria et al. 2012).

On the other hand, the possible mass for the compact object in X-2 can be estimated assuming the source emits at the Eddington limit.
The BH mass of $M_{bh}$ $\sim$ 10 $M_{\sun}$ is found by using the highest luminosity value ($\sim$1.7$\times$10$^{39}$ erg s$^{-1}$) with this assumption. Also considering the dominant contribution from the disk and the disk component model parameter, the inner disk radius can be derived as $R_{in} \sim$ 96$(\cos \theta)^{-0.5}$ km. Then, an upper limit for the compact source mass is estimated as $\leq$ 80 $M_{\sun}$ in X-2 using the inner disk radius (Makishima et al. 2000). Considering this mass range of (10$-$80) $M_{\sun}$, the compact source in X-2 is probably a stellar mass BH ($M \leq 20 M_{\sun}$) or a massive-stellar BH ($20 M_{\sun} \leq M \leq 100M_{\sun}$)(Feng \& Soria 2011).

In addition, we have detected a long periodicity of $(5.2\pm2.0)\times10^{-5}$ Hz ($\sim$5.4 hrs) above 3 $\sigma$ confidence level in the low state of the source. We speculate that this may be another low frequency QPO from the system that may possibly be consistent with an intermediate mass BH scenario owing to the plausible large size disk. In such a case, this may be the lowest QPO detected from a ULX. However, it is also possible that this is the orbital period of the underlying binary which, then, it will be more consistent with a stellar/massive-stellar size BH. We note that the source shows low and high states of luminosity and there is not enough data on the source to conclusively decide whether it belongs to XRB class or a ULX classification.

We encourage further monitoring X-ray observations of X-2 in the nearby galaxy NGC 4736 to understand its true physical parameters 
and characteristics. In addition, these should be aided with observations in the other wavelength ranges (optical, IR, radio), as well.

\acknowledgments

The authors thank an anonymous referee for the critical reading of the manuscript which helped to improve the paper. The authors also thank M. E. Ozel for his very valuable comments. The authors acknowledge support from the Scientific and Technical Research Council of Turkey (TUBITAK) through project No. 113F039.

\end{document}